# Strong and Tunable Spin Lifetime Anisotropy in Dual-Gated Bilayer Graphene

Jinsong Xu, Tiancong Zhu, Yunqiu Kelly Luo, Yuan-Ming Lu, and Roland K. Kawakami

*Department of Physics, The Ohio State University, Columbus, OH 43210, USA*

**ABSTRACT**

We report the discovery of a strong and tunable spin lifetime anisotropy with excellent spin lifetimes up to 7.8 ns in dual-gated bilayer graphene. Remarkably, this realizes the manipulation of spins in graphene by electrically-controlled spin-orbit fields, which is unexpected due to graphene's weak intrinsic spin-orbit coupling. We utilize both the in-plane magnetic field Hanle precession and oblique Hanle precession measurements to directly compare the lifetimes of out-of-plane vs. in-plane spins. We find that near the charge neutrality point, the application of a perpendicular electric field opens a band gap and generates an out-of-plane spin-orbit field that stabilizes out-of-plane spins against spin relaxation, leading to a large spin lifetime anisotropy. This intriguing behavior occurs because of the unique spin-valley coupled band structure of bilayer graphene. Our results demonstrate the potential for highly tunable spintronic devices based on dual-gated 2D materials.

Graphene is an outstanding material for spin transport because its low spin-orbit coupling (SOC) [1-3] leads to record long spin diffusion lengths at room temperature [4-6]. However, such weak SOC makes it difficult to electrically control spins in graphene, which is highly desired for spintronic device operation [7-10]. Therefore, attention has focused on utilizing stacked heterostructures for graphene spintronics [11-29]. For example, experiments have shown that adjacent ferromagnetic insulator layers can strongly modulate spin currents via proximity exchange fields [11-16], while adjacent transition metal dichalcogenide (TMDC) layers enable the optical injection of spin [17,18] and electrical modulation of spin currents via spin absorption [19,20]. More recently, groundbreaking measurements of large spin lifetime anisotropy (comparing lifetimes of out-of-plane vs. in-plane spins) in graphene-TMDC heterostructures have unambiguously identified the presence of proximity-induced SOC and its impact of spin transport in graphene [25,26]. While this holds great promise for the electrical control of spins in graphene, the observed spin lifetimes in these heterostructures are short (<50 ps) due to the relatively strong SOC in TMDCs.

In this Letter, we report the remarkable observation of a strong and tunable spin lifetime anisotropy in dual-gated bilayer graphene (BLG) with spin lifetimes up to 7.8 ns. The strong spin lifetime anisotropy is unexpected due to graphene's weak intrinsic SOC, but occurs because applying a perpendicular electric field breaks the inversion symmetry of Bernal-stacked BLG, which opens a band gap and dramatically alters the character of the SOC [30-33]. Our experiments are performed on BLG lateral spin valves with top and bottom gates to independently control the perpendicular electric field and carrier density. The spin lifetime anisotropy ($\xi = \tau_\perp/\tau_\parallel$ where $\tau_\perp$ ($\tau_\parallel$) is the lifetime for spins oriented out-of-plane (in-plane)) is determined by means of both in-plane magnetic field Hanle and oblique Hanle measurements [25,26]. We observe strong anisotropy ($\xi$ up to 12.2) near the charge neutrality point (CNP) when an electric field is applied. A detailed study shows that $\xi$ first increases with applied electric field, then saturates and eventually decreases. This intriguing behavior can be explained by the unique band structure of BLG [32,33] where an electric field tunes the gap, and the band edges have spin-orbit splitting (tens of μeV) and spin-valley coupling similar to TMDCs [34]. This produces gate-tunable effective spin-orbit fields that protect out-of-plane spins from dephasing by Dyakonov-Perel spin relaxation. These results demonstrate strong and gate-tunable spin lifetime anisotropy with excellent spin lifetime in BLG, which provides a prototype 2D system for next generation spintronic devices based on electrical control of effective spin-orbit fields.

Spin valve devices (Figure 1a) are fabricated by exfoliating and stacking layers of BLG and hexagonal boron nitride (h-BN) onto a Si wafer with 300 nm $SiO_2$ overlayer. For the bottom gate, we utilize the highly p-doped Si wafer as a global backgate with the 300 nm $SiO_2$ serving as the gate dielectric. For the top gate, we utilize a patterned Cr/Au electrode separated from the BLG by an h-BN dielectric layer. For spin transport, non-magnetic Cr/Au electrodes (E1, E4, both ~ 1 kΩ) are used as reference contacts and Co electrodes with SrO tunneling barriers [35] (E2 ~ 22.8 kΩ, E3 ~3.7 kΩ) are used for spin injection and detection. Details of device fabrication are provided in the Supplemental Material (SM) section S1 [36-38].

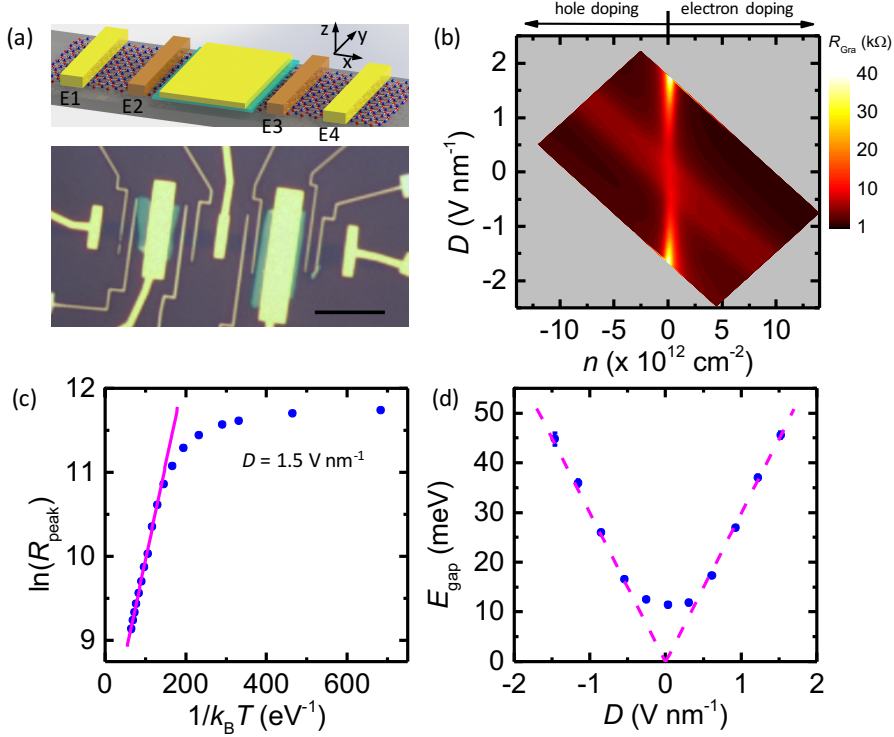

**FIG. 1.** (a) (Top) Schematic of the device. Yellow: Cr/Au electrode, Brown: Co electrode, Cyan: h-BN, Grey: SiO$_2$. (Bottom) Optical image of a BLG spin valve. The scale bar is 10 μm. (b) Electric field and carrier density dependent BLG resistance. (c) Temperature dependent BLG channel peak resistance $R_{peak}$ at $D$ = 1.5 V nm$^{-1}$. The line is an Arrhenius fitting with a slope of 23 meV. (d) Electric field dependent BLG band gap $E_{gap}$. The error bars are the standard errors of the fitted slopes in Figure 1c. The dashed lines indicate the linear relationship between E$_{gap}$ and the electric field, and the deviation near zero field is likely due to resistance contributions unrelated to the gap.

First, we characterize the dependence of BLG channel resistance at 100 K as a function of top gate voltage ($V_t$) and bottom gate voltage ($V_b$) using four-point resistance measurements (current applied between E1 and E4 while measuring the voltage between E2 and E3). The result is plotted in Figure 1b as a function of carrier density ($n$, positive for electrons) and perpendicular electric field ($D$), which are related to $V_b$ and $V_t$, by $n = \frac{\varepsilon_t \varepsilon_0}{d_t e}(V_t - V_{t0}) + \frac{\varepsilon_b \varepsilon_0}{d_b e}(V_b - V_{b0})$ and $D = -\frac{\varepsilon_t}{d_t}(V_t - V_{t0}) + \frac{\varepsilon_b}{d_b}(V_b - V_{b0})$, where $\varepsilon_t$ ($\varepsilon_b$) is the relative dielectric constant of h-BN (SiO$_2$), $\varepsilon_0$ is the vacuum dielectric constant, $d_t$ ($d_b$) is the thickness of h-BN (SiO$_2$), $e$ is the electron charge, $V_{t0}$ ($V_{b0}$) is the effective top (bottom) gate voltage offset due to initial environmental doping. Detailed calculations are shown in SM section S2 [36].

The key features of the data (Figure 1b) are a resistance maximum as a function of $n$, which occurs at the charge neutrality point (CNP) located at $n = 0$ cm$^{-2}$. This resistance maximum, $R_{peak}$, increases with electric field $D$ due to the opening of a band gap. In addition to this feature, there is also a broader resistance maximum that appears as a diagonal ridge. This is due to resistance contributions in the regions of BLG outside of the top gate (thus modulated only by the bottom gate), which is outside of our region of interest. Returning to the main peak features, by measuring the temperature dependence of $R_{peak}$ for a constant value of $D$, we extract $E_{gap}$ by Arrhenius fitting (Figure 1c) [30]. By repeating this for different values of $D$, we obtain the dependence of band gap on electric field (Figure 1d). The maximum $E_{gap}$ is about 50 meV within the electric field applied, which is consistent with previous reports [31].

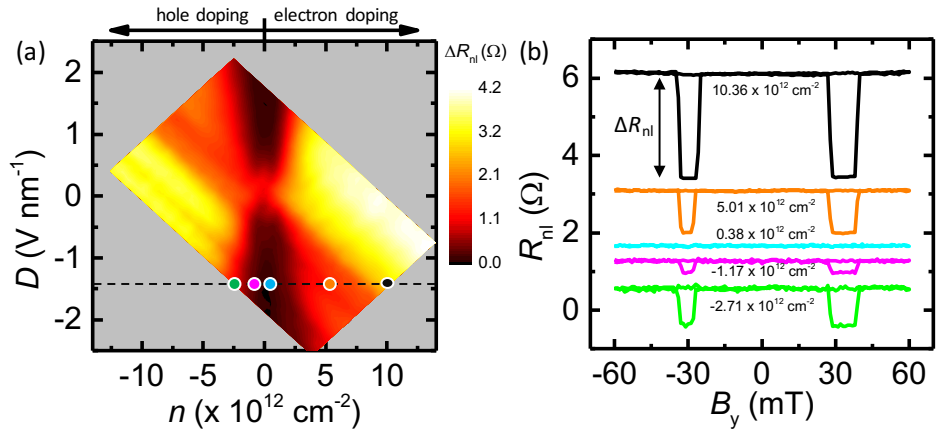

**FIG. 2.** (a) Electric field and carrier density dependence of non-local spin signal $\Delta R_{nl}$. (b) Non-local magnetoresistance $R_{nl}$ as a function of $B_y$ at $D = -1.4$ V nm$^{-1}$ for various carrier densities (dashed line in Figure 2a).

Turning our attention to spin transport, Figure 2 shows the spin transport signal $\Delta R_{nl}$ as a function of $n$ and $D$. This is obtained using the non-local spin transport geometry: a current source $I_{inj}$ applied between the electrodes E1 and E2 (spin injector) creates spins in the BLG beneath E2 which subsequently diffuse toward the spin detector E3, where it is measured as a voltage signal $V_{nl}$ across electrodes E3 and E4. The four-terminal non-local resistance is defined as $R_{nl} = V_{nl}/I_{inj}$. Figure 2b shows a detailed scan of $R_{nl}$ as a function of magnetic field $B_y$ at $D = -1.4$ V nm$^{-1}$ for different $n$, corresponding to the five points in Figure 2a. A hysteretic jump is observed as Co electrode magnetizations switch between parallel (high $R_{nl}$) and antiparallel (low $R_{nl}$) configurations. The presence of these jumps indicates spin transport in the BLG, and the non-local spin signal is defined as $\Delta R_{nl} = R_{nl}$ (parallel) $- R_{nl}$ (antiparallel), as indicated by the arrow in Figure 2b. The detailed dependence of $\Delta R_{nl}$ on $n$ and $D$ is summarized in Figure 2a. The main feature is that $\Delta R_{nl}$ decreases while approaching the CNP and further decreases to zero with an applied $D$.

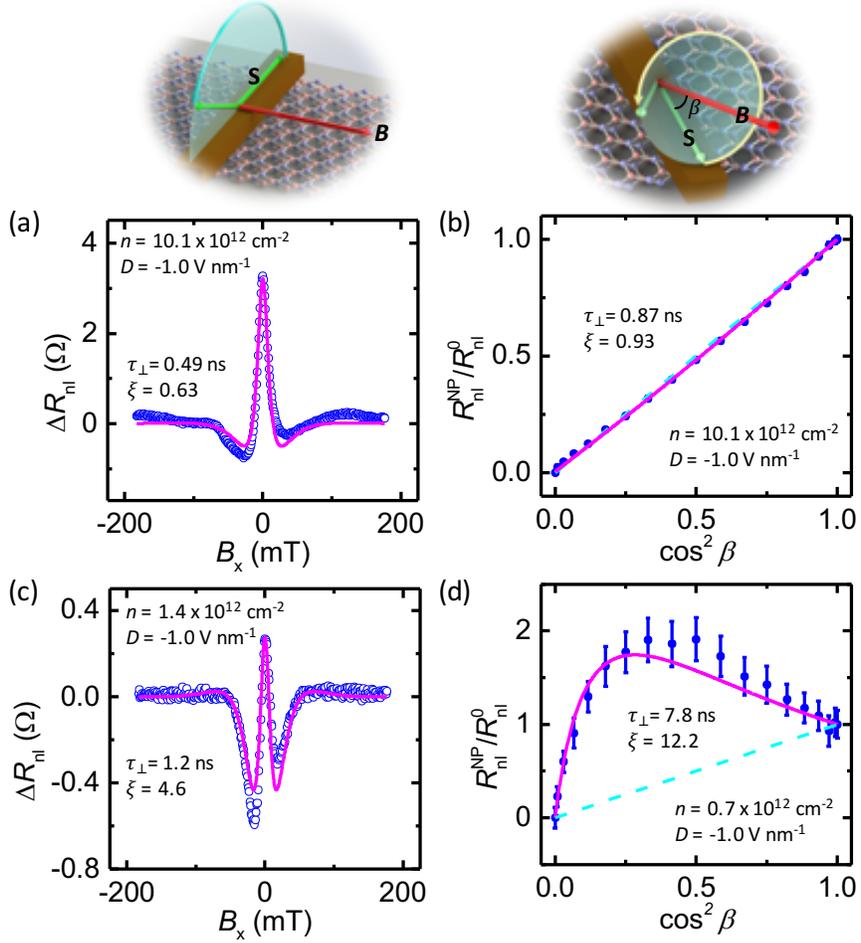

**FIG. 3.** (a) and (c) In-plane magnetic field Hanle at $D$ = -1.0 V nm$^{-1}$ for two different carrier densities. (b) and (d) Oblique Hanle at $D$ = -1.0 V nm$^{-1}$ for two different carrier densities. Blue dots are data, the magenta curve is the fit and the cyan dashed line is the guideline of y = x (no anisotropy). The error bars are standard deviations of the measured data points.

To understand this strong tunability of $\Delta R_{nl}$, we perform both in-plane magnetic field Hanle and oblique Hanle measurements. The in-plane magnetic field Hanle measurement is done by applying a magnetic field $B_x$ along x direction so that spins precess in the y-z plane as it transports across the BLG [25] (see Figure 3a inset). Oblique Hanle measurements are done by applying a magnetic field $\boldsymbol{B} = (B_y, B_z)$ in y-z plane with various angles $\beta$ between magnetic field $\boldsymbol{B}$ and Co electrode so that spins precess about a cone as it transports across the BLG [39-41] (see Figure 3b inset). Because the precession involves both in-plane and out-of-plane spin components, these spin precession measurements can be used to determine the spin lifetime anisotropy $\xi$.

Figures 3a and 3b show the results for in-plane magnetic field Hanle and oblique Hanle, respectively, for high electron density, $n$ =10.1 × 10$^{12}$ cm$^{-2}$. In this regime, the Hanle curves are similar to those observed typically in graphene spin valves without substantial spin lifetime anisotropy [4,39]. For the in-plane magnetic field Hanle (Figure 3a), the curve has a maximum spin signal at zero field which decreases with increasing $B_x$ due to precessional dephasing. The small negative lobes are typical of Hanle curves when the channel length (here $L$ = 7 μm) is longer than the spin diffusion length. For quantitative analysis of the data, we employ a one-dimensional spin diffusion model that allows for different values of in-plane ($\tau_\parallel$) and out-of-plane ($\tau_\perp$) spin lifetime (see SM section S3 for details [36]). The fitting (magneta curve) yields a spin lifetime anisotropy value of $\xi$ = 0.63, with corresponding spin lifetime values of $\tau_\parallel$ = 0.77 ns and $\tau_\perp$ = 0.49 ns. For anisotropy $\xi$ on the order of 1 or less, the oblique Hanle method yields more reliable values. Figure 3b summarizes the key data obtained from a series of oblique Hanle curves to determine $\xi$. Each data point comes from an oblique Hanle curve taken for a different value of $\beta$ (see SM section S4 for details [36]). This plots the non-precessing component of spin signal ($R_{nl}^{NP}$) normalized by the in-plane spin signal ($R_{nl}^0$) as a function of $cos^2\beta$. For $\xi$ = 1, the value of $R_{nl}^{NP}$ will vary linearly with $cos^2\beta$ (specifically, $R_{nl}^{NP}/R_{nl}^0 = cos^2\beta$ ), while the $R_{nl}^{NP}/R_{nl}^0$ vs. $cos^2\beta$ curve will lie *below* the linear curve for $\xi$ < 1, and the $R_{nl}^{NP}/R_{nl}^0$ vs. $cos^2\beta$ curve will lie *above* the linear curve for $\xi$ > 1. Figure 3b shows that the measured $R_{nl}^{NP}/R_{nl}^0$ (blue dots) as a function of $cos^2\beta$ lies slightly below the linear curve (dashed cyan line). A quantitative fit (magneta line) yields a value of $\xi$ = 0.93, with corresponding values of $\tau_\parallel$ = 0.94 ns and $\tau_\perp$ = 0.87 ns. Comparing the results from in-plane magnetic field Hanle ($\xi$ = 0.63) and oblique Hanle ($\xi$ = 0.93), we consider the latter method as more reliable quantitatively, while the former method is more useful for rapidly identifying $\xi$ from a single spin-precession scan.

Moving closer to the CNP, we observe a dramatically different shape for the in-plane magnetic field Hanle curve, as shown in Figure 3c for $n$ = 1.4 x $10^{12}$ cm$^{-2}$ and $D$ = - 1.0 V nm$^{-1}$. Here, the negative lobes become even larger than the peak at zero magnetic field and is very different from any Hanle curve without spin lifetime anisotropy. Such a curve was first reported by Ghiasi *et al.* [25] for graphene-MoSe$_2$ heterostructures and occurs when $\xi \gg 1$. Fitting this curve (magneta line) yields a large spin lifetime anisotropy value of $\xi$ = 4.6, with corresponding spin lifetime values of $\tau_\parallel$ = 0.26 ns and $\tau_\perp$ = 1.20 ns. Figure 3d shows results for oblique Hanle measurements taken for $n$ = 0.7 x $10^{12}$ cm$^{-2}$ and $D$ = -1.0 V nm$^{-1}$. Notably, the $R_{nl}^{NP}/R_{nl}^0$ vs. $cos^2\beta$ curve is substantially higher than the linear curve (dashed cyan line), which is indicative of a large $\xi$. A quantitative fit yields a large spin lifetime anisotropy value of $\xi$ = 12.2, with corresponding spin lifetime values of $\tau_\parallel$ = 0.64 ns and $\tau_\perp$ = 7.8 ns. This strong spin lifetime anisotropy $\xi$ = 12.2 in BLG is comparable to monolayer graphene-TMDC heterostructures [25,26], but with a much longer spin lifetime.

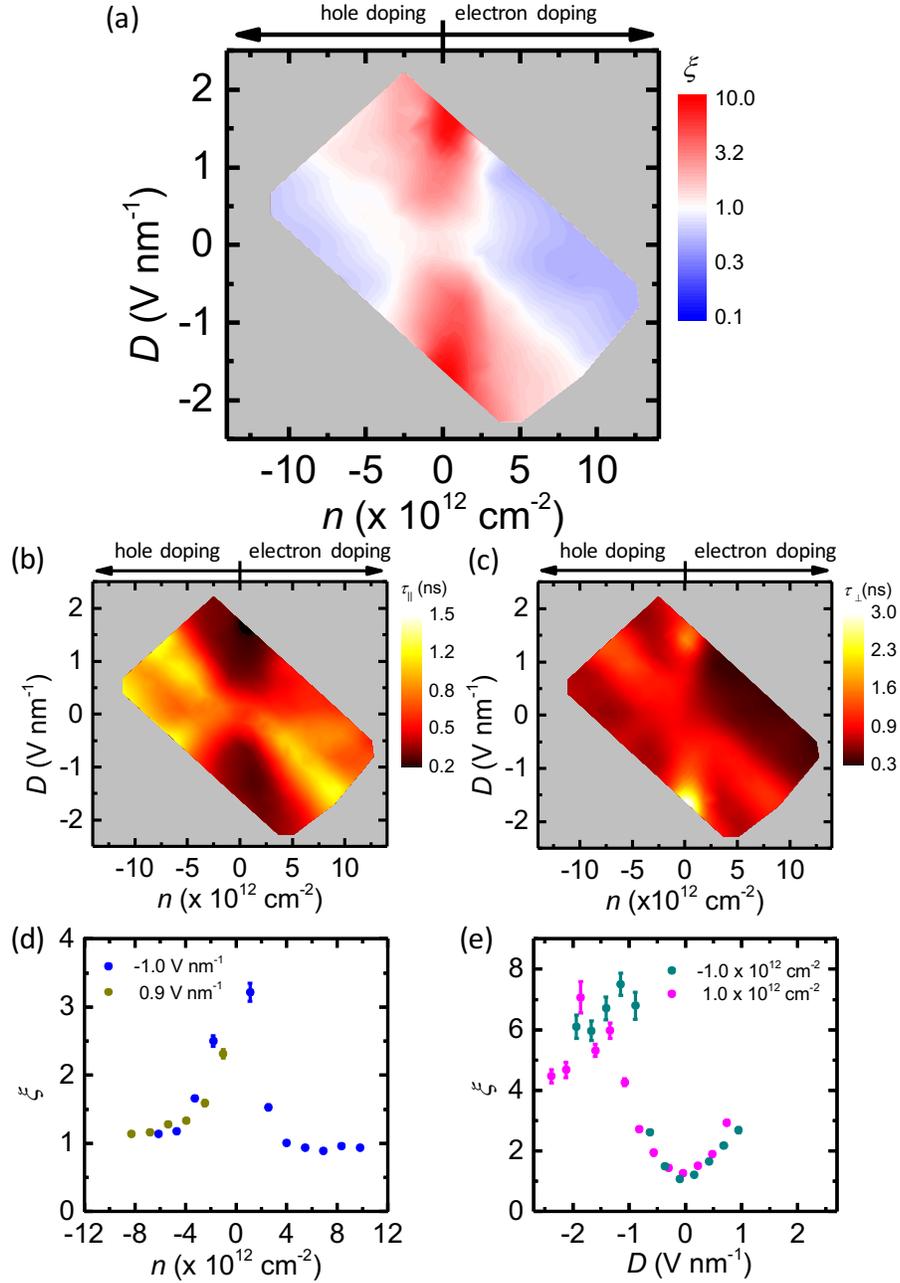

**FIG. 4.** (a-c) Electric field and carrier density dependence of spin lifetime anisotropy, in-plane spin lifetime, and out-of-plane spin lifetime, respectively, extracted from in-plane magnetic field Hanle measurements. (d, e) Spin lifetime anisotropy as a function of carrier density and electric field, respectively, extracted from oblique Hanle measurements.

To understand the detailed role of carrier density and electric field on spin lifetime anisotropy, we utilize in-plane magnetic field Hanle measurements to efficiently generate 2D maps of $\xi$, $\tau_\parallel$ and $\tau_\perp$ as a function of $n$ and $D$ (Figures 4a, 4b, and 4c), and then perform selective line cuts using oblique Hanle measurements to quantitatively assess the variations with $n$ (Figure 4d) and $D$ (Figure 4e). A number of trends are observed. As shown in Figure 4a, it is clear the carrier density needs to be near charge neutrality ($n$ = 0 cm$^{-2}$) and an electric field must be applied in order to have a large $\xi$. To understand whether the large anisotropy is due to an increase of $\tau_\perp$ and/or a decrease of $\tau_\parallel$, we consider the maps in Figures 4b and 4c. Starting at the CNP and $D$ = 0 V nm$^{-1}$, increasing the magnitude of $D$ leads to both a reduction of $\tau_\parallel$ (Figure 4b) and an enhancement of $\tau_\perp$ (Figure 4c). For the $\tau_\parallel$ map, it is interesting to note that this resembles the map of the non-local spin signal in Figure 2a, which makes sense because the non-local spin transport is for spins oriented in-plane. For the $\tau_\perp$ map, we notice a diagonal ridge similar to that observed in Figure 1b. This is also likely to originate from regions of the graphene with only bottom gate and is thus outside the primary region of interest. Fortunately, similar features appear for both $\tau_\parallel$ and $\tau_\perp$ so they have only minor impact on the $\xi$ map. Turning to the oblique Hanle line cuts, the dependence of $\xi$ on carrier density for different fixed electric fields shows a large value near charge neutrality with applied $D$ = -1.0 V nm$^{-1}$ or 0.9 V nm$^{-1}$ and a reduction to $\xi \sim 1$ with increasing carrier density (Figure 4d). Figure 4e shows the dependence of $\xi$ on electric field $D$ for carrier densities $n = -1.0$ x $10^{12}$ cm$^{-2}$ and 1.0 x $10^{12}$ cm$^{-2}$. Toward negative $D$, the $\xi$ first increases with magnitude of electric field, then saturates around 1 V nm$^{-1}$ and eventually decreases. This is observed for both the electron and hole doping. Unfortunately, the trend for positive $D$ could not be tested to similarly high values due to a shift of the gate voltage offsets ($V_{b0}$ and $V_{t0}$) after sample reload [42].

This electric field and carrier density dependence of the spin lifetime anisotropy can be qualitatively understood using a simple phenomenological model. Since the low-energy physics is controlled by the electronic band structure of BLG near zone corner K and K' [32,33], as schematically shown in Figure 5a, the out-of-plane spin lifetime $\tau_\perp$ should increase with the out-of-plane spin polarization $\langle S_Z \rangle$ at K and K'. At finite temperature, we have $\langle S_Z \rangle \propto \pm(-\dfrac{1}{Exp\left[\frac{\frac{E_{gap}}{2}-\mu+2\lambda_I-\mu}{k_B T}\right]+1} + \dfrac{1}{Exp\left[\frac{\frac{E_{gap}}{2}-\mu}{k_B T}\right]+1} + \dfrac{1}{Exp\left[\frac{\frac{E_{gap}}{2}-\mu}{k_B T}\right]+1} - \dfrac{1}{Exp\left[\frac{\frac{E_{gap}}{2}-2\lambda_I-\mu}{k_B T}\right]+1})$, where + (-) corresponds to K (K') valley, $\lambda_I$ is the intrinsic SOC, $\mu$ is the chemical potential (using the middle of the gap as zero potential), and $T$ is temperature. Figure 5b is the plot of out-of-plane spin polarization $\langle S_Z \rangle$, with $E_{gap}$ = 30 $D$ meV ($D$ in units of V nm$^{-1}$) extracted from Figure 1d, $\lambda_I$ = 12 μeV, $T$ = 100 K and $\mu$ = 1 meV. $\langle S_Z \rangle$ first increases with electric field $D$, then saturates and decreases eventually, implying a similar $D$-dependence for the out-of-plane spin lifetime $\tau_\perp$. This toy model without detailed calculation well explains the observed electric field dependence of the spin lifetime anisotropy (Figure 4e). In addition, $\langle S_Z \rangle$ will decay quickly moving away from the K or K' points in momentum space due to rapid decrease of $\lambda_I$, resulting in the decrease of spin lifetime anisotropy with increasing carrier density, as observed experimentally (Figure 4d). These results are consistent with Dyakonov-Perel spin relaxation in the presence of out-of-plane spin-orbit fields, which will reduce the in-plane spin lifetime (Figure 4b) by inducing additional precessional dephasing, while also stabilizing the out-of-plane spins against precessional dephasing for enhanced spin lifetimes (Figure 4c).

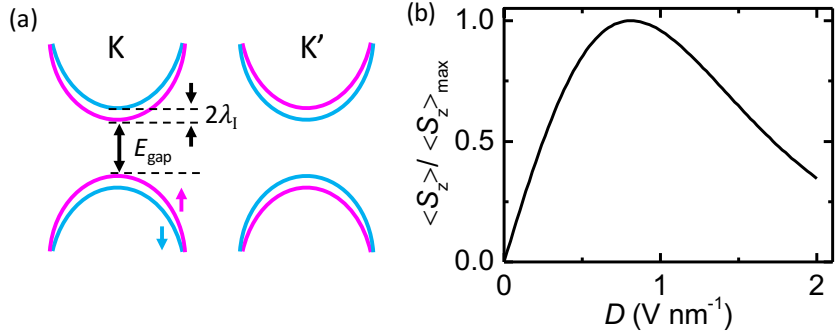

**FIG. 5.** (a) Schematic drawing of the BLG band structure with applied electric field. (b) Electric field dependence of normalized $\langle S_z \rangle$ with $E_{gap}$ = 30 $D$ meV ($D$ in units of V nm$^{-1}$), $\lambda_I$ = 12 $\mu$eV, $T$ = 100 K and $\mu$ = 1 meV.

In conclusion, we have observed large and tunable spin lifetime anisotropy in dual-gated BLG spin valves. Investigating the dependence on electric field and carrier density, we find spin lifetime anisotropy as strong as $\xi$ = 12.2 with an out-of-plane spin lifetime up to 7.8 ns. Near the CNP, the spin lifetime anisotropy first increases with electric field, then saturates and eventually decreases, which is explained using a simple model of the spin-valley coupled band structure of BLG. Our results demonstrate the potential for highly tunable spintronic devices based on dual-gated 2D materials.

*Note.* During preparation of this manuscript, we became aware of a similar result using single-gated BLG [43].

## ACKNOWLEDGEMENTS

We acknowledge technical assistance from Jyoti Katoch, Simranjeet Singh, and Dongying Wang. Funding for this research was provided by the Center for Emergent Materials: an NSF MRSEC under award number DMR-1420451.

Supplemental Material for:

# Strong and Tunable Spin Lifetime Anisotropy in Dual-Gated Bilayer Graphene

Jinsong Xu, Tiancong Zhu, Yunqiu Kelly Luo, Yuan-Ming Lu, and Roland K. Kawakami

*Department of Physics, The Ohio State University, Columbus, OH 43210, USA*

**Section S1. Device fabrication**

A dry transfer technique is used for fabricating BLG spin valves with h-BN top gate. First, we mount ~2 mm thick polydimethylsiloxane (PDMS) on a glass slide and cover it with a thin film of polycarbonate (PC). This PC/PDMS stamp is used to pick up the h-BN flake (10~20 nm) from an SiO₂/Si substrate. This h-BN flake is then aligned and brought into contact with BLG on another SiO₂/Si substrate. After contact, the PC film is cut from the glass slide and the entire PC/h-BN/BLG combination remains on SiO₂/Si substrate. The PC film is then dissolved in chloroform. After that, the transferred h-BN/BLG heterostructure is cleaned of polymer residue by annealing at 350 °C in ultra-high vacuum (UHV) for 1 hour. Then we use two steps of e-beam lithography with MMA/PMMA bilayer resist to fabricate electrodes. In the first step, Au electrodes (70 nm) are deposited on the h-BN/BLG heterostructure using an e-beam source and a 5 nm Cr underlayer for adhesion. In the second step, Co electrodes with SrO tunnel barriers are deposited using angle evaporation with polar angle of 0° for the SrO masking layer (3 nm), 10° for the SrO tunnel barrier (0.8 nm), and 6° for the Co electrode (35 nm) in an MBE chamber.

**Section S2. Determination of carrier density and electric field**

The top and bottom gate dependence of BLG resistance is shown in Figure S1 (a). The resistance exhibits a peak value $R_{peak}$ along the diagonal line, which corresponds to CNP line. In addition, there is a horizontal line where the resistance does not depend on $V_t$, which is due to the contribution of BLG outside the top gate region. This is the same feature seen in the main text Figure 1b. To see this more clearly, Figure S1 (b) shows the bottom gate dependence of BLG resistance at several fixed top gate voltages. As seen, besides the main peak $R_{peak}$, there is a side peak at $V_b = $ -8 V for all top gates, resulting from BLG region outside the top gate.

The carrier density and electric field are calculated using Eq. (1) and (2)

$$n = \frac{\varepsilon_t \varepsilon_0}{d_t e}(V_t - V_{t0}) + \frac{\varepsilon_b \varepsilon_0}{d_b e}(V_b - V_{b0}) \tag{1}$$

$$D = -\frac{\varepsilon_t}{d_t}(V_t - V_{t0}) + \frac{\varepsilon_b}{d_b}(V_b - V_{b0}) \tag{2}$$

The device is fabricated on 300 nm $SiO_2$, therefore $\varepsilon_b = 3.9$, and $d_b = 300$ nm, which gives $\frac{\varepsilon_b}{d_b} = 0.013$ nm$^{-1}$. By tracing out the $R_{peak}$ position ($n = 0$) as a function of $V_t$ and $V_b$, we can calculate the ratio of $(\frac{\varepsilon_t}{d_t})/(\frac{\varepsilon_b}{d_b})$ (Figure S1 (c)). For the device presented in main text, $(\frac{\varepsilon_t}{d_t})/(\frac{\varepsilon_b}{d_b}) = 11.45$, so $\frac{\varepsilon_t}{d_t} = 0.149$ nm$^{-1}$. Because the BLG band gap $E_{gap}$ is proportional to electric field $D$, and $R_{peak}$ increases with $E_{gap}$, is, $V_{t0}$ and $V_{b0}$ can be obtained by finding the $V_t$ and $V_b$ values where $R_{peak}$ reaches a minimum. We find $V_{t0} = -1$ V and $V_{b0} = -8$ V for the device presented in main text. Therefore, for the device presented in main text we have

$$n = 7.194 \times 10^{10} \times [11.45 \times (V_t + 1) + (V_b + 8)] \tag{3}$$

$$D = 0.013 \times [-11.45 \times (V_t + 1) + (V_b + 8)] \tag{4}$$

where $n$ is in units of cm$^{-2}$, $D$ is in units of V nm$^{-1}$ and $V_t$, $V_b$ are in units of V.

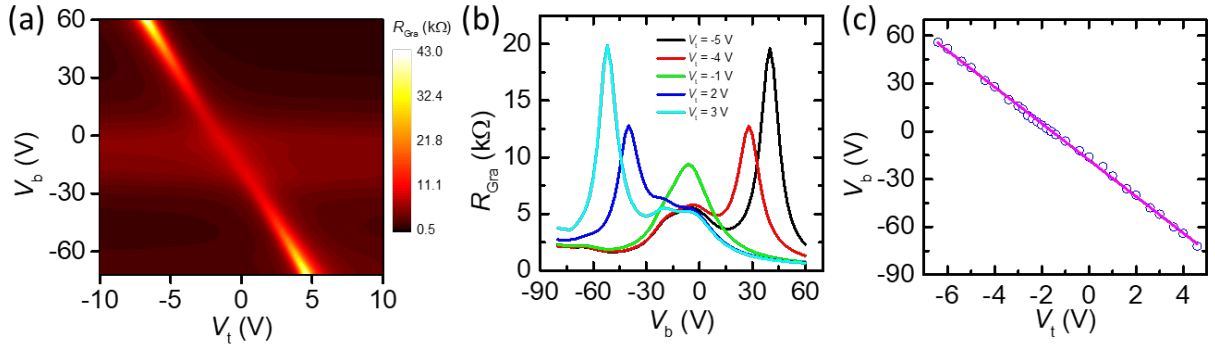

**Figure S1.** (a) Top and bottom gate dependence of BLG resistance. (b) Bottom gate dependence of BLG resistance at several fixed top gate voltages. (c) $V_b$ as a function of $V_t$ for the $R_{peak}$ position.

## Section S3. Spin diffusion model for in-plane magnetic field Hanle and oblique Hanle

For in-plane magnetic field Hanle measurement, $\boldsymbol{B} = B_0\boldsymbol{e}_x$ is applied along the $x$ direction. The Bloch equation can be written as

$$D_s\mu_x'' - \frac{\mu_x}{\tau_\parallel} = 0, \tag{5a}$$

$$D_s \mu_y'' - \frac{\mu_y}{\tau_\parallel} + \omega_0 \mu_z = 0, \tag{5b}$$

$$D_s \mu_z'' - \frac{\mu_z}{\tau_\perp} - \omega_0 \mu_y = 0. \tag{5c}$$

where $D_s$ is the spin diffusion coefficient, $\mu_j$ (j = x, y, z) is the spin accumulation along j-axis direction and $\omega_0 = (g\mu_B/\hbar)B_0$. Following the same procedure in ref [1,2], the spin accumulation is solved as

$$\mu_y(L) = \frac{1}{2\sqrt{\gamma^2 - 1}}[(\gamma + \sqrt{\gamma^2 - 1})\mu_p(L) + (-\gamma + \sqrt{\gamma^2 - 1})\mu_m(L)], \tag{6a}$$

with

$$\mu_p(L) = \frac{2K_+\Delta}{\eta_0\eta_L e^{-K_+L} - (2K_+ - \eta_0)(2K_+ - \eta_L)e^{K_+L}}, \tag{6b}$$

$$\mu_m(L) = \frac{2K_-\Delta}{\eta_0\eta_L e^{-K_-L} - (2K_- - \eta_0)(2K_- - \eta_L)e^{K_-L}}. \tag{6c}$$

where

$$\eta_i^{-1} = -\sigma^G(R_{Fi}^* \frac{1}{W_{Fi}} + R_{Ci}^* \frac{1}{W_{Ci}}), \tag{7a}$$

$$\Delta = -\eta_0 J^C \left(P_\sigma^F R_{F0}^* \frac{W_{C0}}{W_{F0}} + P_\sigma^{C0} R_{C0}^*\right), \tag{7b}$$

$$K_+ = \sqrt{\frac{1}{D_s}\left[\frac{1}{\tau_\parallel} + \omega_0\left(-\gamma + \sqrt{\gamma^2 - 1}\right)\right]}, \tag{7c}$$

$$K_- = \sqrt{\frac{1}{D_s}\left[\frac{1}{\tau_\parallel} - \omega_0\left(\gamma + \sqrt{\gamma^2 - 1}\right)\right]}, \tag{7d}$$

$$\gamma = \frac{1}{2\omega_0\tau_\parallel}\left(1 - \frac{1}{\xi}\right). \tag{7e}$$

$\sigma^G(\sigma^F, \sigma^C)$ is the BLG (Co, contact) conductivity, $R_F^*$ is the effective spin resistance $\frac{\lambda_F}{\sigma^F}\frac{1}{1 - P_\sigma^{F2}}$ of Co, $R_C^*$ is the effective spin resistance $\frac{1}{\sigma^C}\frac{1}{1 - P_\sigma^{C2}}$ of the contact, $P_\sigma^F$ ($P_\sigma^C$) is the conductance polarization of Co (contact), $W_F$ ($W_C$) is the width of the Co (contact) and $J^C$ is the current density passing contact. And the non-local MR is

$$R_{nl} = -P_0 P_L \frac{R_G}{L} \frac{\mu_y(L)}{\Delta}$$

$$= -P_0 P_L \frac{R_G}{L} \left[ \frac{\left(\frac{\gamma}{\sqrt{\gamma^2-1}}+1\right)K_+}{\eta_0 \eta_L e^{-K_+ L} - (2K_+ - \eta_0)(2K_+ - \eta_L)e^{K_+ L}} + \right.$$

(8a)

$$\left. \frac{\left(-\frac{\gamma}{\sqrt{\gamma^2-1}}+1\right)K_-}{\eta_0 \eta_L e^{-K_- L} - (2K_- - \eta_0)(2K_- - \eta_L)e^{K_- L}} \right],$$

$$P_0 = \frac{P_\sigma^0 R_{C0}^* + P_\sigma^F R_{F0}^* \frac{W_{C0}}{W_{F0}}}{R_{C0}^* + R_{F0}^* \frac{W_{C0}}{W_{F0}}}, \tag{8b}$$

$$P_L = \frac{P_\sigma^L R_{CL}^* + P_\sigma^F R_{FL}^* \frac{W_{CL}}{W_{FL}}}{R_{CL}^* + R_{FL}^* \frac{W_{CL}}{W_{FL}}}. \tag{8c}$$

All in-plane magnetic field Hanle data in this paper was fit using Eq.(8a).

For the oblique Hanle measurement, instead of applying $\boldsymbol{B}$ field along the $x$ direction, we apply $\boldsymbol{B}$ field in the $y-z$ plane, and the angle between $+y$ and $\boldsymbol{B}$ is $\beta$, i.e.

$$\boldsymbol{B} = (0, B_0 cos\beta, B_0 sin\beta), \tag{9a}$$

$$\boldsymbol{\omega} = (0, \omega_0 cos\beta, \omega_0 sin\beta). \tag{9b}$$

In order to solve the Bloch equation, following in ref [3], we transform it into a new coordinate system, of which the $y'$ direction is along $\boldsymbol{B}$ field direction, $x'$ is the same as $x$ and $z'$ changes correspondingly. The transformation matrix is

$$T = \begin{pmatrix} 1 & 0 & 0 \\ 0 & cos\beta & sin\beta \\ 0 & -sin\beta & cos\beta \end{pmatrix}. \tag{10}$$

In this new coordinate system, we use $s$ to represent the spin accumulation $\mu$ for clarity, then the Bloch equation can be rewritten as

$$D_s s_x'' - \frac{s_x}{\tau_\parallel} - \omega_0 s_z = 0, \tag{11a}$$

$$D_s s_y'' - \left(\frac{cos^2\beta}{\tau_\parallel} + \frac{sin^2\beta}{\tau_\perp}\right)s_y - \left(\frac{-sin\beta cos\beta}{\tau_\parallel} + \frac{sin\beta cos\beta}{\tau_\perp}\right)s_z = 0, \tag{11b}$$

$$D_s s_z'' - \left(\frac{-sin\beta cos\beta}{\tau_\parallel} + \frac{sin\beta cos\beta}{\tau_\perp}\right)s_y - \left(\frac{sin^2\beta}{\tau_\parallel} + \frac{cos^2\beta}{\tau_\perp}\right)s_z + \omega_0 s_x = 0. \tag{11c}$$

In the limit $\omega_0 \rightarrow \infty$ $(B_0 \rightarrow \infty)$, $s_x$ and $s_z$ are dephased much faster than $s_y$, Eqs. (11) become

$$D_s s_x'' - \omega_0 s_z = 0, \tag{12a}$$

$$D_s s_y'' - \frac{1}{\tau_\parallel}\left(cos^2\beta + \frac{sin^2\beta}{\xi}\right)s_y = 0, \tag{12b}$$

$$D_s s_z'' + \omega_0 s_x = 0. \tag{12c}$$

Following the same procedure for solving Bloch equation for the in-plane magnetic field Hanle measurement, we have the spin accumulation as

$$s_y(L) = -\Delta_1 \lambda_\parallel f_1 = -\Delta \lambda_\parallel f_1 cos\beta, \tag{13a}$$

$$s_z(L) = -\Delta_2 \lambda_\parallel f_2 = \Delta \lambda_\parallel f_2 sin\beta, \tag{13b}$$

$$f_1 = \left\{ 2\left[ \sqrt{cos^2\beta + \frac{sin^2\beta}{\xi}} + \frac{\lambda_\parallel}{2}\left(\frac{1}{r_0} + \frac{1}{r_L}\right)\right] e^{\left(\frac{L}{\lambda_\parallel}\right)\sqrt{cos^2\beta + \frac{sin^2\beta}{\xi}}} \right.$$

$$\left. + \frac{\lambda_\parallel^2}{r_0 r_L} \frac{sinh\left[\left(\frac{L}{\lambda_\parallel}\right)\sqrt{cos^2\beta + \frac{sin^2\beta}{\xi}}\right]}{\sqrt{cos^2\beta + \frac{sin^2\beta}{\xi}}} \right\}^{-1}, \tag{13c}$$

$$f_2 = Re\left\{2\left[\sqrt{-i\omega_0\tau_\parallel} + \frac{\lambda_\parallel}{2}\left(\frac{1}{r_0} + \frac{1}{r_L}\right)\right]e^{\left(\frac{L}{\lambda_\parallel}\right)\sqrt{-i\omega_0\tau_\parallel}} + \right.$$

$$\left. \frac{\lambda_\parallel^2}{r_0 r_L}\frac{\sinh\left[\left(\frac{L}{\lambda_\parallel}\right)\sqrt{-i\omega_0\tau_\parallel}\right]}{\sqrt{-i\omega_0\tau_\parallel}}\right\}^{-1}. \tag{13d}$$

The non-local MR is

$$R_{\text{nl}}^{\text{NP}}(\beta) = -P_0 P_L \frac{R_G}{L}\frac{s_y(L)cos\beta - s_z(L)sin\beta}{\Delta} \tag{14}$$

$$= P_0 P_L R_N(f_1(\beta)\cos^2\beta + f_2(\beta)\sin^2\beta).$$

All oblique Hanle data in this paper was fit using Eq. (14) to determine $\xi$.

### Section S4. Oblique Hanle measurement

To perform the oblique Hanle measurement in main text, we measure the non-local spin signal $R_{\text{nl}}$ at $B = 180$ mT with various angles $\beta$. And we also do detailed Hanle measurements for various angle $\beta$ to spot check that the spin component perpendicular to $\boldsymbol{B}$ fully dephases. Figure S2 (a) and (b) are examples of detailed Hanle measurements for spin lifetime anisotropy near unity and greater than 1, respectively. The blue dots in Figure S2 (c) and (d) are the summarized $R_{\text{nl}}^{\text{NP}}/R_{\text{nl}}^0$ by averaging $R_{\text{nl}}$ data between $B = 160$ mT and 180 mT in Figure S2 (a) and (b). The error bar corresponds to the standard deviation of $R_{\text{nl}}$ between $B = 160$ mT and 180 mT. The magenta data is the $R_{\text{nl}}^{\text{NP}}/R_{\text{nl}}^0$ by directly measuring $R_{\text{nl}}$ at $B = 180$ mT, and the error bar corresponds to standard deviation of $R_{\text{nl}}$ at $B = 180$ mT for multiple measurements. The comparison of the two measurement methods (blue and magneta dots) show that they are essentially the same.

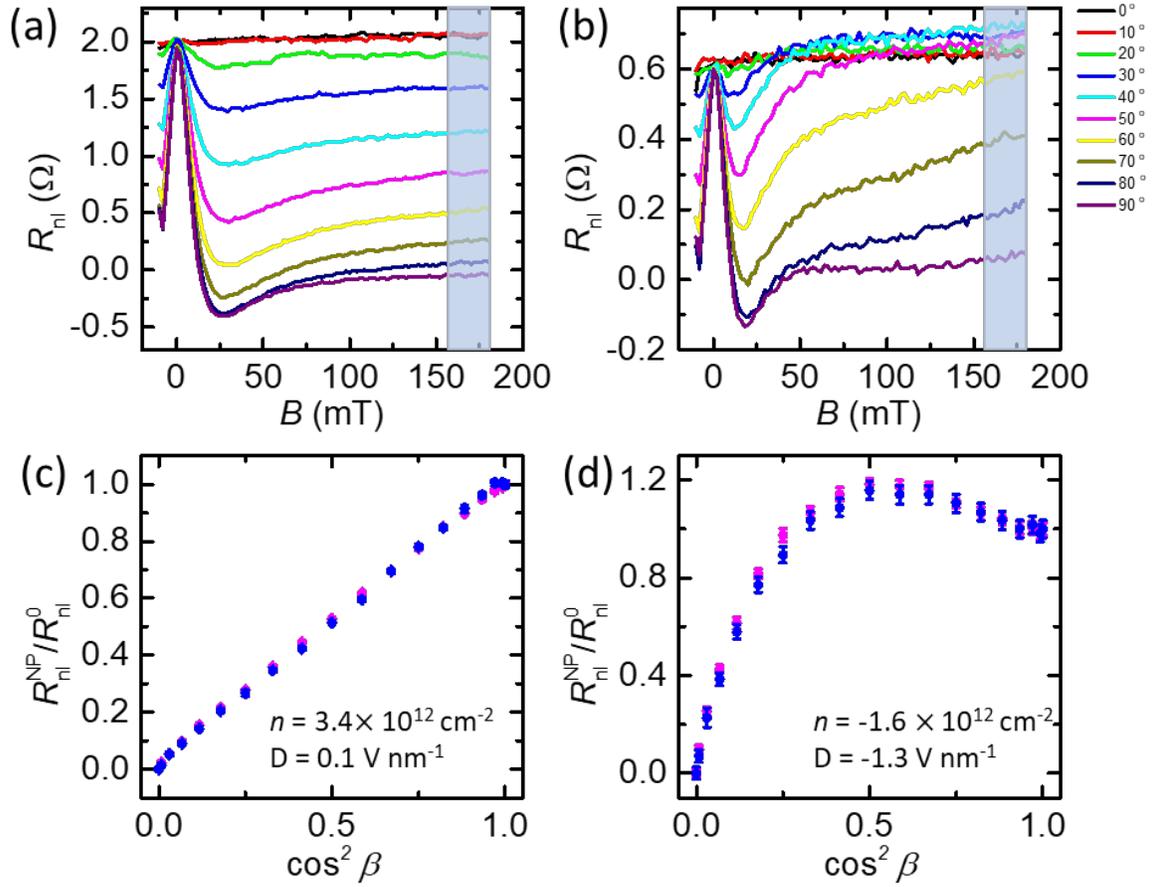

**Figure S2.** Detailed Hanle measurement at various angles $\beta$ for (a) $n = 3.4 \times 10^{12}$ cm$^{-2}$, $D = 0.1$ V nm$^{-1}$ and (b) $n = -1.6 \times 10^{12}$ cm$^{-2}$, $D = -1.3$ V nm$^{-1}$. (c) and (d) are the corresponding $R_{nl}^{NP}/R_{nl}^{0}$. The blue data the average of $R_{nl}^{NP}/R_{nl}^{0}$ between $B = 160$ mT and 180 mT from (a) and (b). The magenta data are $R_{nl}^{NP}/R_{nl}^{0}$ measured directly at 180 mT for multiple measurements.